\documentclass[aps,twocolumn,prl,superscriptaddress,showpacs,tightenlines]{revtex4}
\usepackage{amssymb}
\usepackage{amsmath}
\usepackage{graphicx}
\usepackage{epsfig}
\usepackage{subfigure}
\usepackage{amsfonts}

\begin{document}

\title{Controllable Scattering of a single photon inside a one-dimensional resonator waveguide}
\author{Lan Zhou}
\affiliation{Advanced Science Institute, The Institute of Physical
and Chemical Research (RIKEN), Wako-shi 351-0198, Japan}
\affiliation{Department of Physics, Hunan Normal University,
Changsha 410081, China}
\author{Z. R. Gong}
\affiliation{Institute of Theoretical Physics, The Chinese Academy
of Sciences, Beijing, 100080, China}
\author{Yu-xi Liu}
\affiliation{Advanced Science Institute, The Institute of Physical
and Chemical Research (RIKEN), Wako-shi 351-0198, Japan}
\affiliation{CREST, Japan Science and Technology Agency (JST),
Kawaguchi, Saitama 332-0012, Japan}
\author{C. P. Sun}
\affiliation{Institute of Theoretical Physics, The Chinese Academy
of Sciences, Beijing, 100080, China}
\author{Franco Nori}
\affiliation{Advanced Science Institute, The Institute of Physical
and Chemical Research (RIKEN), Wako-shi 351-0198, Japan}
\affiliation{CREST, Japan Science and Technology Agency (JST),
Kawaguchi, Saitama 332-0012, Japan} \affiliation{Center for
Theoretical Physics, Physics Department, Center for the Study of
Complex Systems, The University of Michigan, Ann Arbor, MI
48109-1040, USA.}

\begin{abstract}

We analyze the coherent transport of a single photon, which
propagates in a one-dimensional coupled-resonator waveguide and is
scattered by a controllable two-level system located inside one of
the resonators of this waveguide. Our approach, which uses discrete
coordinates, unifies ``low'' and ``high'' energy effective theories
for single-photon scattering. We show that the controllable
two-level system can behave as a quantum switch for the coherent
transport of a single photon. This study may inspire new
electro-optical single-photon quantum devices. We also suggest an
experimental setup based on superconducting transmission line
resonators and qubits.
\end{abstract}

 \maketitle

\emph{Introduction.---} The scattering of a structureless particle
can be used to determine the internal structure of a scattering
target. This has been well recognized since the Rutherfold
experiments which ushered modern particle and nuclear
physics~\cite{landau}. When scattering is confined to low
dimensions, it displays new features. For example, the low-energy
scattering of cold atoms confined in an atomic waveguide can form a
gas of impenetrable bosons exhibiting total reflection~\cite{tonks}.
Such total reflection, and related phenomena, motivate us to study
low-dimensional photonic scattering, oriented towards quantum
information processing. Specifically: how to control the coherent
transport of scattered single photon by tuning the inner structure
of the target so that the target can behave as a quantum switch;
i.e., either a perfect mirror totally reflecting photons, or an
ideal transparent medium allowing photons to pass. Based on
theoretical studies of photonic scattering in one dimensional(1D)
waveguides~\cite{fanpaper}, an all-optical single-photon transistor
was recently proposed~\cite{Lukin-np} by using surface plasmons
confined in a conducting nano-wire.

Here we study a quantum switch that controls the transport of a
confined single-photon. The switch is a scattering target made of a
controllable two-level system. Our approach recovers the interesting
results obtained via an effective field theory~\cite{fanpaper} in
the ``high'' energy regime. Furthermore, it can also be consistently
applied to the ``low'' energy regime discussed below. We show that
the total reflection by a controllable two-level system can be
realized as a resonant-scattering phenomenon and the reflection
spectrum goes beyond the Breit-Wigner~\cite{landau} and Fano
lineshapes~\cite{fano,arkady}.

As an application of our study, we propose an
experimentally-accessible quantum electro-optical device,
constructed using superconducting transmission line
resonators~\cite{jqyou,cdfqho,yale1} and a superconducting charge
qubit~\cite{YouPT58,GWe0508,nature1}. In our proposed device, the
scattering target is a charge qubit with two energy levels
controlled by a gate voltage and an external magnetic flux; the
coupled-transmission-line resonators behave as a 1D continuum for
the coherent transport of photons. Thus, the controllable charge
qubit can be used to manipulate the coherent transport of photons in
an array of superconducting transmission line resonators.

\begin{figure}[tbp]
\includegraphics[bb=100 310 520 560, width=7 cm,clip]{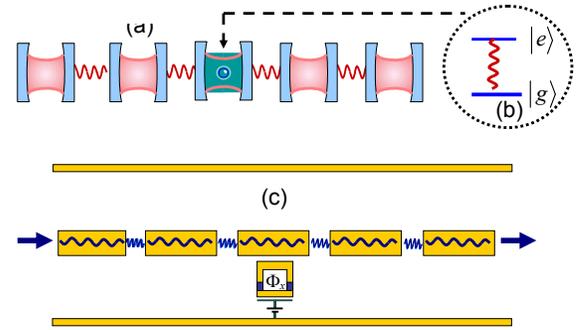}
\caption{(Color online). Schematic configuration for the coherent
transport of a single-photon in a coupled resonator waveguide (a)
coupled to a two-level system (b), which is located in one of the
resonators. (c) Schematic diagram of coupled superconducting
transmission line resonators with one resonator coupled to a
dc-SQUID-based charge qubit.} \label{fig:1}
\end{figure}

\emph{Discrete scattering equation.---}We consider a 1D
coupled-resonator~\cite{FNRMP} waveguide (CRW) (see
Fig.~\ref{fig:1}) with a two-level system, which is embedded in
one of the resonators. The CRW can be realized by using either
coupled superconducting transmission line resonators~\cite{zhgsun}
or defect resonators in photonic crystals~\cite{greentree}. Let
$a_{j}^{\dagger}$ ($ j=-\infty,\cdots, \infty$) be the creation
operator of the $j$th single mode cavity with frequency $\omega$.
The Hamiltonian for the CRW is given by
\begin{equation}
H_{c}=\omega \sum_{j}a_{j}^{\dagger }a_{j}-\xi \sum_{j}\left(
a_{j}^{\dagger }a_{j+1}+{\rm h.c.}\right)   \label{lps-01}
\end{equation}%
with the inter-cavity coupling constant $\xi$, which describes
photon hopping from one cavity to another. Here, we assume that all
resonators have the same frequency $\omega$ and $\hbar=1$. The
Hamiltonian~(\ref{lps-01}) describes a typical tight-binding boson
model, which has the dispersion relation $\Omega _{k}=\omega
-V_{k}$, with $V_{k}=2\xi\cos(l k)$. Below, the lattice constant $l$
is assumed to be unity. In the low-energy regime, corresponding to
long wavelengths ($\lambda\gg l$), the spectrum is quadratic:
$\Omega _{k}\simeq \omega _{\xi }+\xi k^{2}$, with $\omega _{\xi
}=\omega -2\xi $. At the matching condition ($\lambda\sim 4l$), the
spectrum is linear: $\Omega _{k}\simeq \omega _{\xi }\pm2\xi k$. In
contrast to the similar configurations in
Refs.~\cite{zhjsun76,hu76PRA,Plenio,Sergey}, here only one two-level
system, with ground state $\left\vert g\right\rangle $, excited
state $\left\vert e\right\rangle $ and transition energy $\Omega$,
is located inside one of coupled cavities. Moreover,
Refs.~\cite{zhjsun76,hu76PRA,Plenio} use several approximations
which we do not make here, making the treatment here more physical,
since we have exact solutions. For convenience, we take the $0$th
cavity as the coordinate-axis origin and we also assume that a
two-level system is located in this $0$th cavity~\cite{note}. Under
the rotating wave approximation, the interaction between the $0$th
cavity field and the two-level system is described by a
Jaynes-Cummings Hamiltonian
\begin{equation}
H_{I}=\Omega \left\vert e\right\rangle \left\langle e\right\vert
+J\left( a_{0}^{\dag }\left\vert g\right\rangle \left\langle
e\right\vert +\left\vert e\right\rangle \left\langle g\right\vert
a_{0}\right),   \label{lps-02}
\end{equation}
with the coupling strength $J$.

To study the 1D single-photon elastic scattering described by the
total Hamiltonian $H=H_{c}+H_{I}$, we assume the stationary
eigenstate
\begin{equation}
|\Omega_{k}\rangle =\sum_{j}u_{k}\left( j\right) a_{j}^{\dag
}\left\vert 0\rangle\vert g\right\rangle +u_{e}\left\vert
0\rangle\vert e\right\rangle, \label{lps-03}
\end{equation}%
when a single photon comes from the left with eigenenergy
$\Omega_{k}$.  Here, $\left\vert 0\right\rangle$ is the vacuum state
of the cavity field, and $u_{e}$ is the probability amplitude of the
two-level system in the excited state. This $H|\Omega _{k}\rangle
=\Omega _{k}|\Omega _{k}\rangle $ results in the discrete scattering
equation
\begin{equation}
\left( V_{k}+JG_{k}\delta _{j0}\right) u_{k}\left( j\right) =\xi
\left[ u_{k}(j+1)+u_{k}\left( j-1\right) \right].  \label{lps-04}
\end{equation}%
Here, the Green function $G_{k}=G_{k}(\Omega )=J/\left(
\Omega_{k}-\Omega \right)$, and $u_{e}=G_{k}u_{k}\left( 0\right)$
relates the excited state amplitude $u_{e}$ with the single-photon
amplitude $u_{k}\left( 0\right)$.

\emph{Reflection and transmission amplitudes.---}
Equation~(\ref{lps-03}) presents a complete set of stationary states
of the total system for single-photon processes. The scattering
equation $V_{k}u_{k}\left( j\right) =\xi \left[ u_{k}\left(
j+1\right) +u_{k}\left( j-1\right) \right]$ for $j\neq 0$ has the
solution
\begin{equation}
u_{k}\left( j\right) =\left\{
\begin{array}{c}
u_{Lk}\left( j\right) =e^{ikj}+re^{-ikj}\text{, }j<0 \\
u_{Rk}\left( j\right) =se^{ikj}\text{, \ \ \ \ \ \ }j>0%
\end{array}%
\right.   \label{lps-06}
\end{equation}%
with transmission and reflection amplitudes $s$ and $r$. The
continuous condition $u_{k}\left( 0^{+}\right) =u_{k}\left(
0^{-}\right)$ and the eigenvalue equation $\left(
V_{k}+JG_{k}\right) u_{k}\left( 0\right) =\xi \left[ u_{k}\left(
1\right) +u_{k}\left( -1\right) \right]$ at $j=0$ determine the
reflection amplitude
\begin{equation}
r=J^{2}\left[2i\xi \sin k\left( \omega -\Omega -2\xi \cos k\right)
-J^{2}\right]^{-1} \label{lps-08}
\end{equation}%
and the transmission amplitude $s$, with the constraints $s=r+1$
and $|s|^{2}+|r|^{2}=1$.
\begin{figure}[tbp]
\includegraphics[width=8 cm]{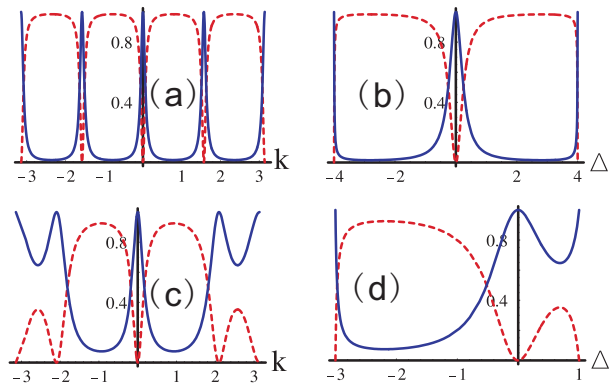}
\caption{(Color online). The reflection coefficient $R$ (blue solid
line) and the transmission coefficient, $1-R$ (red dashed line) as a
function of either the momentum $k$ or the detuning
$\Delta=\Omega_{k}-\Omega$, when $\omega =\Omega =5$ with
intercavity coupling $\xi =2$, [for (a,b)]; and $\omega =5,\Omega
=6$ with $\xi =1$ [for (c,d)]. Parameters are in units of $J$.}
\label{fig:2}
\end{figure}

As shown below, Eq.~(\ref{lps-08}) is a very useful result.
Figs.~\ref{fig:2}(a,c) show the reflection coefficient
$R(k)=|r(k)|^{2}$ versus the momentum $k$ of the incident photons,
while Figs.~\ref{fig:2}(b,d) plot the reflection coefficient
$R(\Delta )=|r(\Delta)|^{2}$ versus the detuning $\Delta
=\Omega_{k}-\Omega $. Figure~\ref{fig:2}(b) represents a
Breit-Wigner-like lineshape around the resonance $\Delta =0$, where
the line-width is proportional to $J^{2}$. At the resonance, the
photon is completely reflected and the single two-level system
behaves as a \textit{perfect mirror}. Therefore, when the two-level
system has a tunable transition energy, it can be used as \textit{a
quantum switch to control the coherent transport of photons}.

\begin{figure}[tbp]
\includegraphics[width=8 cm]{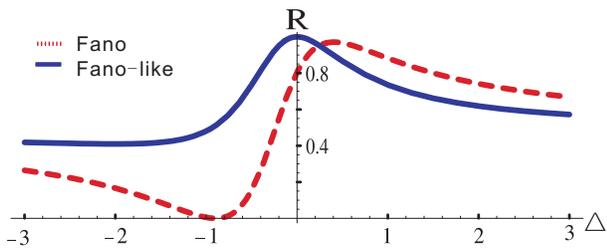}
\caption{(Color online). Comparison between the Fano-like lineshape
(blue solid line) in Eq.~(\ref{eq:9}) and the Fano lineshape (red
dashed line) given by $F\left( \Delta \right) =\left( \Delta
/\protect\delta _{\protect\xi }+y_{0}-q\Gamma
\right)^{2}\left[\left(\Delta /\delta _{\xi
}+y_{0}\right)^{2}+\Gamma ^{2}\right]^{-1}$ at $\delta_{\xi} =-3$
and $\xi=0.01$, where all the parameters are in units of $J$. The
detunings are $\Delta=\Omega_{k}-\Omega$ and
$\delta_{\xi}=\omega-\Omega-2\xi$.} \label{fig:1-1}
\end{figure}

Due to the nonlinear dispersion relation $\Omega_{k}=\omega-2\xi\cos
k$, $|r(k)|^{2}$ in Figs.~\ref{fig:2}(a,c) shows a more general
lineshape, \textit{beyond} the Breit-Wigner~\cite{landau} and
Fano~\cite{fano} lineshapes. Indeed, Eq.~(\ref{lps-08}) is very
general as it can directly provide results in different physical
limits. When $\left\vert \Delta \right\vert <\left\vert \delta _{\xi
}\right\vert$, as shown in Fig.~(\ref{fig:1-1}), the generalized
Fano-like lineshape of the reflection spectrum
\begin{equation}
R\left( \Delta \right)=\frac{\eta \left( 1+\Delta /\delta_{\xi
}+\Delta ^{2}/\delta_{\xi }^{2}\right)}{\left( \Delta /\delta _{\xi
}+y_{0}\right)^{2}+\Gamma ^{2}}\label{eq:9}
\end{equation}%
is approximately obtained from Eq.~(\ref{lps-08}). Here, $\eta
=J^{4}\left(J^{4}-4\xi\delta_{\xi }^{3}\right)^{-1}$, the detuning
$\delta _{\xi }=\omega-\Omega-2\xi $, $y_{0}=\eta /2$, and $\Gamma
^{2}=\eta(1-\eta/4)$.

When $\lambda\sim 4l$, the Breit-Wigner lineshape~\cite{landau}
\begin{equation}
r=-iJ^{2}\left[2\xi \left( \omega_{\pi} -\Omega \pm 2\xi k\right)
+iJ^{2}\right]^{-1} \label{lps-09}
\end{equation}%
is also straightforwardly obtained from Eq.~(\ref{lps-08}) by
expanding the cosine around $k=\pm \pi/2$, where
$\omega_{\pi}=\omega-\pi\xi$. Eq.~(\ref{lps-09}) was also derived
from the continuous field theory in Ref.~\cite{fanpaper}. Indeed, in
this limit ($\lambda\sim 4l$), the Hamiltonian $H_{c}$ for the CRW
can be approximated by $ H_{c}=\sum_{j}(\omega_{\pi}\pm 2\xi
k)\;a_{k}^{\dagger }a_{k}$, producing a linear dispersion relation
as in Ref.~\cite{fanpaper}.

\emph{Low-energy effective theory.---}Let us now consider the
long-wavelength regime ($\lambda\gg l$) and use the low-energy
effective theory to consistently describe the scattering of the
confined photons. In this regime, $k$ is so small that $\cos k\simeq
1-k^{2}/2$, and $\sin k\simeq k$, thus Eq.~(\ref{lps-08}) becomes
\begin{equation}
r\simeq -iJ^{2}\left[2k\xi \left(\xi k^{2}+\omega _{\xi }-\Omega
\right) +iJ^{2}\right]^{-1}\text{.}  \label{lps-10}
\end{equation}

The result in Eq.~(\ref{lps-10}), obtained from Eq.~(\ref{lps-08}),
can also be explained using the low-energy effective field theory
described by the effective Hamiltonian
\begin{eqnarray}
H &=&\int_{-\infty }^{\infty }dx\varphi ^{\dag }\left( x\right) \left(
\omega _{\xi }-\xi \partial _{x}^{2}\right) \varphi \left( x\right) +\Omega
\left\vert e\right\rangle \left\langle e\right\vert   \notag \label{lps-11} \\
&+&\int_{-\infty }^{\infty }dxJ\delta \left( x\right) \left[
\varphi ^{\dag }\left( x\right) \left\vert g\right\rangle
\left\langle e\right\vert +h.c\right].
\end{eqnarray}
Here, the field operator $\varphi \left( x\right) \equiv
\int_{-\infty }^{\infty }dk\exp \left( ikx\right) a_{k}$ satisfies
the commutation relation $\left[ \varphi \left( x\right) ,\varphi
^{\dag }\left( x^{\prime }\right) \right] =\delta \left( x-x^{\prime
}\right) $. Eq.~(\ref{lps-11}) can be derived from the momentum
space representation of the total Hamiltonian
\begin{equation}
H=\sum_{k=0}^{N-1}\Omega _{k}\hat{a}_{k}^{\dag }\hat{a}_{k}+\Omega
\left\vert e\right\rangle \left\langle e\right\vert
+\sum_{k=0}^{N-1}\left( \frac{J\hat{a}_{k}^{\dag
}}{\sqrt{N}}\left\vert g\right\rangle \left\langle e\right\vert
+{\rm h.c.}\right)   \label{lps-13}
\end{equation}%
with $\Omega _{k}=\omega _{\xi }+\xi k^{2}/2$.

We now study single-photon scattering by a two-level system using
the effective field theory in Eq.~(\ref{lps-11}). Let us consider
one photon, with energy $\Omega _{k}$, incident from the left. The
elastic scattering analysis assumes the stationary state $\left\vert
\Omega _{k}\right\rangle =\int_{-\infty }^{\infty }dx\,u_{k}\left(
x\right) \varphi ^{\dag }\left( x\right) \left\vert 0\rangle\vert
g\right\rangle +u_{e}\left\vert 0\rangle\vert e\right\rangle$. Here,
the conservation of total excitation $\sum_{k}n_{k}+\left\vert
e\right\rangle \left\langle e\right\vert =1$ is used. The
eigen-equation $H\left\vert \Omega _{k}\right\rangle =\Omega
_{k}\left\vert \Omega _{k}\right\rangle $ provides a system of
equations for the single-photon probability amplitudes $u_{k}\left(
x\right) $ and the excited-state population $u_{e}$, which results
in a scattering equation with resonance pole
\begin{equation}
\xi \partial _{x}^{2}u_{k}\left( x\right) =G_{k}\delta \left(
x\right) u_{k}\left( 0\right) +\left( \omega _{\xi }-\Omega
_{k}\right) u_{k}\left( x\right).   \label{lps-15}
\end{equation}%

Then the photon scattering can be described by Eq.~(\ref{lps-15}),
which can be solved by assuming $u_{k}\left( x\right) =\exp \left(
ikx\right) +r\exp \left( -ikx\right) $, for $x<0$, and $u_{k}\left(
x\right) =s\exp \left( ikx\right)$, for $x>0$. The reflection
amplitude in Eq.~(\ref{lps-10}) can be obtained from
Eq.~(\ref{lps-15}) with the boundary condition due to the $\delta
$-function
\begin{equation}
\xi \left[ \frac{\partial }{\partial x}u_{k}\left( \epsilon \right) -\frac{%
\partial }{\partial x}u_{k}\left( -\epsilon \right) \right] =\frac{%
J^{2}u_{k}\left( 0\right) }{\Omega _{k}-\Omega }  \label{lps-16}
\end{equation}%
and the continuity $u_{k}\left(\epsilon\right)=u_{k}\left( -\epsilon
\right)$.

\emph{Physical implementation.---} To demonstrate our theoretical
results on the reflection and transmission lineshapes beyond the
Breit-Wigner and Fano lineshapes, we now propose an experimentally
accessible quantum device shown in Fig.~\ref{fig:1}(c), that uses
superconducting transmission line resonators and a dc-SQUID-based
charge qubit~\cite{jqyou,yale1,nature1}. Here, the charge qubit acts
as the scattering target, and its internal structure (e.g., the
transition frequency) is controllable by both the voltage applied to
the gate and the external flux through the SQUID loop. The coupled
coplanar transmission line resonators, constructed by cutting the
superconducting transmission line into $N$ equal
segments~\cite{zhgsun}, provide the continuum for the coherent
transport of photons. The coupling $\xi$ between two neighboring
transmission line resonators is realized via dielectric materials
(it depends on the concrete coupling mechanism). Then the
Hamiltonian of the coupled resonator waveguide is the same as that
in Eq.~(\ref{lps-01}).

The energy eigenstates of the charge qubit~\cite{YouPT58,GWe0508}
are defined by $\left\vert e\right\rangle =\cos (\theta/2)|0\rangle
-\sin (\theta /2)|1\rangle $ and $\left\vert g\right\rangle =\sin
(\theta /2)|0\rangle +\cos (\theta /2)|1\rangle $, with the
transition frequency $\Omega =\sqrt{B_{z}^{2}+B_{x}^{2}}$ for
$\theta =\arctan \left( B_{x}/B_{z}\right)$. $\left\vert
0\right\rangle $ and $\left\vert 1\right\rangle$ are charge
eigenstates representing the excess Cooper pairs on the
superconducting island. The parameter $B_{z}=4E_{C}\left(
2n_{g}-1\right)$, with the charge energy $E_{C}=e^2/2(C_{\rm
g}+2C_{\rm J})$ and $n_{g}=C_{g}V_{g}/2e$, can be controlled by the
voltage $V_{g}$ applied to the gate capacitance $C_{g}$. Here
$C_{\rm J}$ is the capacitance of the Josephson junction. The
parameter $B_{x}=2E_{J}\cos \left( \pi \Phi _{x}/\Phi _{0}\right)$,
with the Josephson energy $E_{J}$, can be changed by the external
magnetic flux $\Phi _{x}$ through the SQUID loop.

As in Ref.~\cite{yale1}, we assume that the charge qubit is placed
in the antinode of the single-mode quantized electric field in the
transmission line resonator with length $L$. Therefore, the
quantized voltage $V_{q}=(\hat{a}+\hat{a}^{\dag })\sqrt{\omega
/(Lc)}$, induced by the quantized electric field, is also applied to
the charge qubit via the gate capacitance $C_{\rm g}$. Here $\omega
$ is the frequency of the quantized field, $c$ is the capacitance
per unit length of the transmission line. The coupling strength
$J=e\sin \theta C_{g}/C_{\Sigma }\sqrt{\omega /(Lc)}$ between the
qubit and the resonator, with $C_{\Sigma }=C_{\rm g}+2C_{\rm J}$,
has feasible values in the range 5--200~MHz~\cite{yale1}. The
detuning $\delta =\omega -\Omega$ between the charge qubit and the
single-mode field can be changed from $-10$~GHz to $10$~GHz. The
frequency of each resonator is in the range between 5--10~GHz, and
the qubit frequency can be tuned from $5$ to $15$~GHz~\cite{yale1}.

\begin{figure}[tbp]
\includegraphics[bb=29 300 570 550, width=9.0cm, clip]{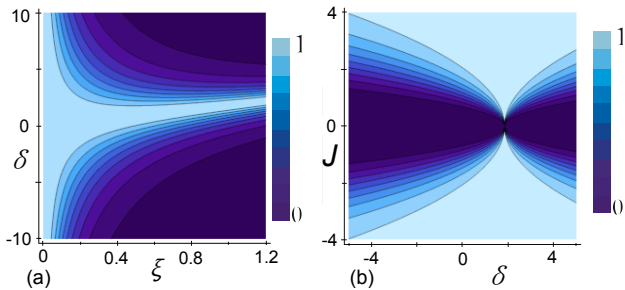}
\caption{(Color online). Phase diagrams of the reflection spectrum
$R(\Delta)$ in Eq.~(\ref{eq:9}): (a) with respect to the intercavity
coupling $\xi$ and the detuning $\delta=\omega-\Omega$, in units of
$J$; (b) with respect to $J $ and $\delta$, in units of $\xi$; both
at $k=\protect\pi/8$.} \label{fig:5}
\end{figure}

Using Eq.~(\ref{lps-08}) and also the parameters given above, we can
show that the scattering process of single-photon degenerates to a
total reflection when the coupling strength $\xi$ between the
resonators vanishes or the incident photon resonates with the qubit.
Also the stronger coupling between the qubit and the resonator
corresponds to the larger reflection amplitude. The details of this
phenomenon are depicted by the contour map of the reflection
coefficient in Fig.~\ref{fig:5}. It can be regarded as a kind of
phase diagram. In the white areas, the reflection is nearly one and
the transmission is almost zero, while in the dark areas, the
transmission approaches unity. The results obtained in this work are
very different from previous quantum switches~\cite{Sun1}.

\emph{Conclusions.---}We have studied the coherent transport of a
single-photon confined in a 1D cavity array. The scattering target
is a controllable two-level system. In the matching regime
($\lambda\sim 4 l$), our approach recovers the
results~\cite{fanpaper} obtained from its effective field theory.
Our approach also predicts a general spectral structure in which the
reflection and the transmission are beyond the usual Breit-Wigner
and Fano lineshapes. These results could be verified experimentally
via a circuit QED system~\cite{jqyou,cdfqho,yale1}. However, in
reality, all large quantum systems interact with the environment,
resulting in some inelastic scattering of photons. Thus the
environment could affect the photon reflection coefficient, reducing
the quantum switching efficiency. The environment-induced inelastic
scattering is related to (i) the decoherence of the resonators; and
(ii) the decay of the two-level system. Case (i) influences the free
propagation of the single photon. The coherent scattering process
happens only when the photon decay rate is much smaller than the
inter-cavity coupling. Case (ii) broadens the width of the lineshape
at the resonance. Finally, we also note that the properties of the
delivered photons at the end of the waveguide could be studied
experimentally by measuring the transmission spectrum of the
resonator.

This work is supported by  NSFC No.~90203018, No.~10474104,
No.~60433050, and No.~10704023, NFRPC No.~2006CB921205 and
2005CB724508. FN acknowledges partial support from the NSA, LPS,
ARO, NSF grant No.~EIA-0130383, JSPS-RFBR 06-02-91200, and the
JSPS-CTC program.


\begin{thebibliography}{99}
\bibitem{note} If the coordinate-axis origin is not taken at the center
of the cavity, where the two-level system is located, the reflection
amplitude in Eq.~(\ref{lps-08}) includes a position-dependent global
phase.
\end{thebibliography}
\end{document}